# Worldline Green Functions for Multiloop Diagrams

Michael G. Schmidt [*]

*Theoretical Physics Division, CERN*
*CH – 1211 Geneva 23*

and

*Institut für Theoretische Physik* [†]
*Universität Heidelberg*
*Philosophenweg 16*
*69120 Heidelberg*

Christian Schubert [‡]

*Institut für Hochenergiephysik Zeuthen*
*DESY Deutsches Elektronen-Synchrotron*
*Platanenallee 6*
*15738 Zeuthen*

## Abstract

We propose a multiloop generalization of the Bern-Kosower formalism, based on Strassler's approach of evaluating worldline path integrals by worldline Green functions. Those Green functions are explicitly constructed for the basic two-loop graph, and for a loop with an arbitrary number of propagator insertions. For scalar and abelian gauge theories, the resulting integral representations allow to combine whole classes of Feynman diagrams into compact expressions.

---

[*] e-mail address k22@vm.urz.uni-heidelberg.de
[†] on sabbatical leave
[‡] e-mail address schubert@hades.ifh.desy

It has always been clear that, in some sense, string theory reduces to particle theory in the limit of infinite string tension. At the one-loop level, this notion was finally made precise by Bern and Kosower [1], who proved that amplitudes in ordinary quantum field theory may be represented as the infinite string tension limit of certain superstring amplitudes. This enabled them to derive a new set of Feynman rules for one-loop calculations in field theory which, though very different from the usual ones, can be seen to be equivalent [2]. Due to the superior organization of string amplitudes – both with respect to gauge invariance and exchange symmetry between the scattering states – these rules lead to a significant reduction of the number of terms to be computed in gauge theory calculations, a fact which has been successfully exploited for the calculation of both N-point gluon-amplitudes [3] and four-point graviton-amplitudes [4].

A simpler derivation of the same rules was given later by Strassler [5], following an approach which had, in a somewhat different context, already been suggested by Polyakov [6]. Here one uses the known representations of one-loop effective actions in terms of superparticle path integrals [7, 8, 9, 10], and treats those path integrals as one-dimensional analogues of the Polyakov integral (more exactly, of the Fradkin-Tseytlin path integral [11]). This approach turned out to be well-suited to the calculation of 1-loop effective actions in general [12, 13], and highly efficient for the calculation of their inverse mass expansions [14].

It would be obviously desirable to generalize this calculus beyond the one-loop level. In the original approach of Bern and Kosower this would imply finding the particle theory limits of higher genus string amplitudes, a formidable task considering the complicated structure of moduli space for $g > 1$. While some work has been done along these lines for the pure gauge theory case [15], it has, as to date, not yet led to the derivation of multiloop rules.

A very different attempt has been made by Lam [16, 17], who started from the usual Feynman parameter integral representation of multiloop diagrams, and used the electric circuit analogy [18] to transform them into the Koba-Nielsen type representation expected from a stringy calculation.

In this letter, we will follow a third route, and show that Strassler's approach can be used to derive a multiloop generalization in a quite elementary way. We will present here just the main line of the argument; the details and some sample calculations will be given in a forthcoming publication [19].

First let us shortly review how one–loop calculations are done in this formalism. For example, for calculation of the one-loop effective action induced by a massive spinor loop in an (abelian or nonabelian) background gauge field plus some scalar (matrix) potential one would start with the worldline

## path integral representation

$$\Gamma[A,V] = -2\int_0^\infty \frac{dT}{T} e^{-m^2 T}\operatorname{tr}\int \mathcal{D}x\mathcal{D}\psi$$
$$\times \exp\Big[-\int_0^T d\tau\Big(\frac{1}{4}\dot x^2 + \frac{1}{2}\psi\dot\psi + igA_\mu\dot x^\mu - ig\psi^\mu F_{\mu\nu}\psi^\nu + V(x)\Big)\Big], \tag{1}$$

where the $x^\mu(\tau)$'s are the periodic functions from the circle with circumference $T$ into $D$ – dimensional Euclidean spacetime, and the $\psi^\mu(\tau)$'s their antiperiodic supersymmetric partners. In the nonabelian case, path ordering is implied.

For calculation of the effective action, one may now split off from the bosonic path integral the integration over the center of mass $x_0$,

$$\int \mathcal{D}x = \int dx_0 \int \mathcal{D}y$$
$$x^\mu(\tau) = x_0^\mu + y^\mu(\tau)$$
$$\int_0^T d\tau\, y^\mu(\tau) = 0, \tag{2}$$

Taylor-expand the external fields at $x_0$, and evaluate the path integral over the relative coordinate $y$ like in string theory, i. e. by Wick contractions in the one-dimensional worldline field theory on the circle. The Green functions to be used are those adapted to the (anti-) periodicity conditions,

$$\langle y^\mu(\tau_1) y^\nu(\tau_2)\rangle = -g^{\mu\nu} G_B(\tau_1,\tau_2) = -g^{\mu\nu}\Big[|\tau_1-\tau_2| - \frac{(\tau_1-\tau_2)^2}{T}\Big],$$
$$\langle \psi^\mu(\tau_1)\psi^\nu(\tau_2)\rangle = \frac{1}{2}g^{\mu\nu} G_F(\tau_1,\tau_2) = \frac{1}{2}g^{\mu\nu}\operatorname{sign}(\tau_1-\tau_2), \tag{3}$$

and the normalization is such that the free path integrals yield

$$\int \mathcal{D}y\, \exp\Big[-\int_0^T d\tau \frac{1}{4}\dot y^2\Big] = [4\pi T]^{-\frac{D}{2}}$$
$$\int \mathcal{D}\psi\, \exp\Big[-\int_0^T d\tau \frac{1}{2}\psi\dot\psi\Big] = 1 \tag{4}$$

For the calculation of 1-loop scattering amplitudes, one would define integrated vertex operators for the external states, e.g. $\int_0^T d\tau\, \exp[ikx(\tau)]$ for scalars or $\int_0^T d\tau\, [\dot x^\mu\varepsilon_\mu - 2\psi^\mu\psi^\nu k_\mu\varepsilon_\nu]\exp[ikx(\tau)]$ for photons, and calculate multiple insertions of vertex operators into the free path integrals. Those written above may be derived from (1) by choosing a plane wave background. For the Dirac particle, a worldline superfield formalism is also available [20, 6]. Yukawa couplings may be introduced, too, but require additional path integrations.

A priori, it is not clear that this formalism should have a useful generalization to higher orders. While worldline path integral representations do exist for multiloop amplitudes (for QED this has even been shown to all orders [21]), this is less clear for worldline Green functions, as the one-loop case is rather special, for the following two reasons:

i) Unlike the circle, a general graph is no manifold, and the Laplacian not well-defined at the node points.

ii) A general graph has no natural global parametrization.

But let us be optimists, and see what we can do.

Consider first some massive scalar field theory, say $\frac{\lambda}{3!}\phi^3$ for simplicity. The usual Feynman-parameter calculation of the one-loop n-point diagram (fig. 1) after performance of the momentum integrations yields an $\alpha$ – parameter integral (we choose n = 4 just for definiteness, and will generally omit coupling constant factors)

$$\int_0^\infty dT \, [4\pi T]^{-\frac{D}{2}} e^{-m^2 T} \prod_{i=1}^{4} \int d\alpha_i \, \delta(T - \alpha_1 - \alpha_2 - \alpha_3 - \alpha_4) \exp[-Q(p_i, \alpha_i)], \quad (5)$$

where

$$TQ = \alpha_2\alpha_4(p_1+p_2)^2 + \alpha_1\alpha_3(p_2+p_3)^2 + \alpha_1\alpha_4 p_1^2 + \alpha_1\alpha_2 p_2^2 + \alpha_2\alpha_3 p_3^2 + \alpha_3\alpha_4 p_4^2 \quad (6)$$

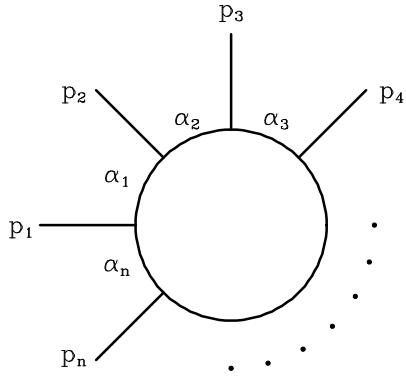

Figure 1: One-loop n-point diagram

In the string-inspired calculation, one would instead calculate the Wick contraction of $n$ tachyon vertex operators

$$\int_0^T d\tau_i \, \exp[ip_i x(\tau_i)] \qquad i = 1,\ldots,n \tag{7}$$

on the circle, using

$$\langle \exp[ip_i x(\tau_i)] \exp[ip_j x(\tau_j)] \rangle = \exp[G_B(\tau_i,\tau_j)p_i p_j], \tag{8}$$

and obtain

$$\int_0^\infty \frac{dT}{T}[4\pi T]^{-\frac{D}{2}} e^{-m^2 T} \frac{1}{n} T \int_0^{\tau_1 = T} d\tau_2 \int_0^{\tau_2} d\tau_3 \cdots \int_0^{\tau_{n-1}} d\tau_n \, \exp\Big[\sum_{i<j} G_B(\tau_i,\tau_j)p_i p_j\Big]. \tag{9}$$

(one integration has been eliminated by using the freedom to choose the zero somewhere on the loop).

Those two integral representations may be transformed into each other by the transformation of variables

$$\begin{aligned} \alpha_1 &= \tau_1 - \tau_2 \\ \cdots & \quad \cdots \\ \alpha_{n-1} &= \tau_{n-1} - \tau_n \\ \alpha_n &= T - (\tau_1 - \tau_n) \end{aligned} \tag{10}$$

and use of momentum conservation. In the abelian case, the stringy formula allows to combine this diagram with those arising by permutations of the external legs by letting all $\tau$ – variables run over the circle independently, yielding [6]

$$\int_0^\infty \frac{dT}{T}[4\pi T]^{-\frac{D}{2}} e^{-m^2 T} \prod_{i=1}^n \int_0^T d\tau_i \, \exp\Big[\sum_{k<l} G_B(\tau_k,\tau_l)p_k p_l\Big]. \tag{11}$$

Let us now look at the naked two-loop graph in the same theory, parametrized as a loop of (proper time) length $T$ with an insertion of length $\bar{T}$ (fig. 2a): As an obvious ansatz for a generalization of the free bosonic path integral to this case, consider the two coupled path integrals

$$\int_0^\infty \frac{dT}{T}\int_0^\infty d\bar{T} e^{-m^2(T+\bar{T})} \int_0^T d\tau_a \int_0^T d\tau_b$$
$$\times \int_{x(0)=x(T)} \mathcal{D}x \, \exp\Big[-\int_0^T d\tau \frac{\dot{x}^2}{4}\Big] \int_{\substack{\bar{x}(0)=x(\tau_a) \\ \bar{x}(\bar{T})=x(\tau_b)}} \mathcal{D}\bar{x} \, \exp\Big[-\int_0^{\bar{T}} d\bar{\tau} \frac{\dot{\bar{x}}^2}{4}\Big] \tag{12}$$

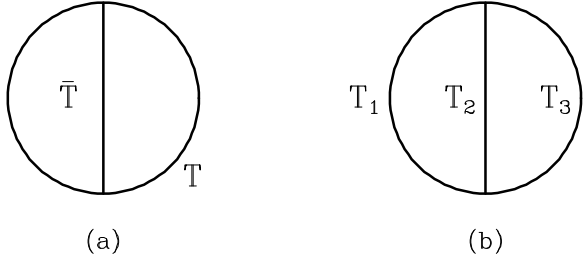

Figure 2: Two different parametrizations of the two-loop diagram

We specialize further, and assume, for the moment, that vertex operators will be inserted only into the $\mathcal{D}x$ – path integral. We may then eliminate the $\mathcal{D}\bar{x}$ – integral – which is, of course, just the path integral representation of the scalar propagator at fixed proper time $\bar{T}$ – to obtain

$$\int_{\substack{\bar{x}(0)=x(\tau_a)\\ \bar{x}(\bar{T})=x(\tau_b)}} \mathcal{D}\bar{x}\ \exp\left[-\int_0^{\bar{T}} d\bar{\tau} \frac{\dot{\bar{x}}^2}{4}\right] = (4\pi\bar{T})^{-\frac{D}{2}} exp\left[-\frac{(x(\tau_a) - x(\tau_b))^2}{4\bar{T}}\right] \quad (13)$$

and remain with

$$\int_{x(0)=x(T)} \mathcal{D}x\ \exp\left[-\int_0^T d\tau \left(\frac{\dot{x}^2}{4} + \frac{(x(\tau_a) - x(\tau_b))^2}{4\bar{T}}\right)\right] \quad (14)$$

Now the additional term in the exponential is quadratic in $x$, so we may hope to absorb it into the free worldline Green function. In fact, its presence corresponds to changing the defining equation for $G_B$ from

$$\int_0^T d\tau_2 \frac{G_B(\tau_1, \tau_2)}{2} \ddot{y}(\tau_2) = y(\tau_1) \quad (15)$$

into

$$\int_0^T d\tau_2 \frac{G_B^{(1)}(\tau_1, \tau_2)}{2} \left[\ddot{y}(\tau_2) - \frac{1}{\bar{T}}[y(\tau_a) - y(\tau_b)][\delta(\tau_2 - \tau_a) - \delta(\tau_2 - \tau_b)]\right] = y(\tau_1) \quad (16)$$

This modified equation can still be solved exactly, yielding

$$G_B^{(1)}(\tau_1, \tau_2) = G_B(\tau_1, \tau_2) + \frac{1}{2} \frac{[G_B(\tau_1, \tau_a) - G_B(\tau_1, \tau_b)][G_B(\tau_2, \tau_a) - G_B(\tau_2, \tau_b)]}{\bar{T} + G_B(\tau_a, \tau_b)} \quad (17)$$

The Green function for calculating insertions into the $\mathcal{D}x$ – integral is thus simply the one-loop Green function plus one additional piece, which incorporates the effect of the internal line.

To determine the normalization of the two-loop path integral compared to the one-loop integral, we have also to calculate the naked Gaussian integral (14), which gives

$$\left[4\pi T + 4\pi \frac{T}{\bar{T}} G_B(\tau_a, \tau_b)\right]^{-\frac{D}{2}} = (4\pi T)^{-\frac{D}{2}} \left[1 + \frac{1}{\bar{T}} G_B(\tau_a, \tau_b)\right]^{-\frac{D}{2}} \qquad (18)$$

Putting things together, our two-loop generalization of eq. (11) becomes

$$\int_0^\infty \frac{dT}{T} \int_0^\infty d\bar{T} e^{-m^2(T+\bar{T})} [4\pi]^{-D} \int_0^T d\tau_a \int_0^T d\tau_b \, [T\bar{T} + T G_B(\tau_a, \tau_b)]^{-\frac{D}{2}}$$
$$\times \prod_{i=1}^n \int_0^T d\tau_i \, \exp\left[\sum_{k<l} G_B^{(1)}(\tau_k, \tau_l) p_k p_l + \frac{1}{2} \sum_{k=1}^n G_B^{(1)}(\tau_k, \tau_k) p_k^2\right] \qquad (19)$$

For fixed $n$, this integral represents the sum of all two-loop diagrams in $\phi^3$ – theory with $n$ legs on the loop, and no leg on the internal line (fig. 3):

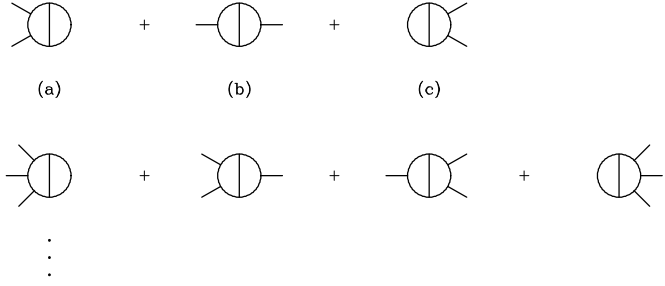

Figure 3: Summation of diagrams with n legs on the loop

Observe that now we have also diagonal terms in the exponential, due to the fact that, in contrast to $G_B$, $G_B^{(1)}$ has $G_B^{(1)}(\tau, \tau) \neq 0$ in general, leading to self-contractions of vertex operators. Those may, if one so wishes, be absorbed into the non-diagonal ones by introducing the "renormalized" Green function

$$\tilde{G}_B^{(1)}(\tau_1, \tau_2) = G_B^{(1)}(\tau_1, \tau_2) - \frac{1}{2} G_B^{(1)}(\tau_1, \tau_1) - \frac{1}{2} G_B^{(1)}(\tau_2, \tau_2) \qquad (20)$$

(this follows immediately from momentum conservation).

For $n = 2, 3$ we have checked this result by transforming it into the Feynman parameter representation. For instance, calculating graph (a) of fig. 3 by Feynman parameters results in

$$\int_0^\infty d\hat{T} \, [4\pi]^{-D} e^{-m^2 \hat{T}} \prod_{i=1}^5 \int d\alpha_i \, \delta(\hat{T} - \sum_{i=1}^5 \alpha_i) \, [P^{(a)}(\alpha_i)]^{-\frac{D}{2}} \exp[-Q^{(a)}(\alpha_i) k^2], \qquad (21)$$

with

$$\begin{aligned} P^{(a)} &= \alpha_5(\alpha_1 + \alpha_2 + \alpha_3 + \alpha_4) \\ P^{(a)} Q^{(a)} &= \alpha_1[\alpha_5(\alpha_2 + \alpha_3 + \alpha_4) + \alpha_2\alpha_3 + \alpha_3\alpha_4] \end{aligned} \qquad (22)$$

The analogue of the transformation eq. ( 10) can be directly read off fig. 4a

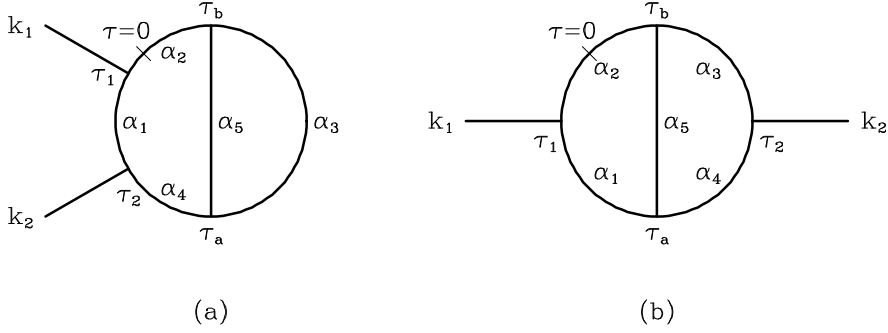

Figure 4: Reparametrization of the two-point two-loop diagrams

$$\begin{aligned} \alpha_1 &= \tau_1 - \tau_2 \\ \alpha_2 &= T - \tau_1 + \tau_b \\ \alpha_3 &= \tau_a - \tau_b \\ \alpha_4 &= \tau_2 - \tau_a \\ \alpha_5 &= \bar{T} \end{aligned} \tag{23}$$

and it does indeed lead to the identifications

$$\begin{aligned} P^{(a)} &= T\bar{T}[1 + \frac{1}{\bar{T}}G_B(\tau_a, \tau_b)] \\ Q^{(a)} &= \tilde{G}_B^{(1)}(\tau_1, \tau_2). \end{aligned} \tag{24}$$

The Feynman calculation for diagram (b) yields polynomials

$$\begin{aligned} P^{(b)} &= \alpha_5(\alpha_1 + \alpha_2 + \alpha_3 + \alpha_4) + (\alpha_1 + \alpha_2)(\alpha_3 + \alpha_4) \\ P^{(b)}Q^{(b)} &= \alpha_5(\alpha_2 + \alpha_3)(\alpha_1 + \alpha_4) + \alpha_1\alpha_2(\alpha_3 + \alpha_4) + \alpha_3\alpha_4(\alpha_1 + \alpha_2) \end{aligned} \tag{25}$$

which are different as functions of the variables $\alpha_i$, but after the corresponding transformation

$$\begin{aligned} \alpha_1 &= \tau_1 - \tau_a \\ \alpha_2 &= \tau_b + T - \tau_1 \\ \alpha_3 &= \tau_2 - \tau_b \\ \alpha_4 &= \tau_a - \tau_2 \\ \alpha_5 &= \bar{T} \end{aligned} \tag{26}$$

identify with the *same* expressions (24) (it should be noted that this becomes apparent only after everything has been expressed in terms of $G_B$, as the absolute signs contained in that function work differently in both cases).

The three $\alpha$ – parameter integrals arising in the calculation of diagrams (a), (b) and (c) thus do indeed correspond to different sectors of our $\tau$ – parameter integral (19).

The whole procedure generalizes without difficulty to the case of $m$ propagator insertions, resulting in an integral representation combining into one expression all diagrams with $n$ legs on the loop, and $m$ inserted propagators:

$$\int_0^\infty \frac{dT}{T} T^{-\frac{D}{2}} [4\pi]^{-(m+1)\frac{D}{2}} \prod_{j=1}^m \int_0^\infty d\bar{T}_j e^{-m^2(T+\sum_{j=1}^m \bar{T}_j)} \int_0^T d\tau_{a_j} \int_0^T d\tau_{b_j}$$
$$\times \prod_{i=1}^n \int_0^T d\tau_i \, N^{(m)\frac{D}{2}} \exp\Big[\frac{1}{2}\sum_{k,l=1}^n G_B^{(m)}(\tau_k,\tau_l) p_k p_l\Big], \qquad (27)$$

where

$$N^{(m)} = \mathrm{Det}(A)$$
$$G_B^{(m)}(\tau_1,\tau_2) = G_B(\tau_1,\tau_2)$$
$$+ \frac{1}{2}\sum_{k,l=1}^m [G_B(\tau_1,\tau_{a_k}) - G_B(\tau_1,\tau_{b_k})] A_{kl} [G_B(\tau_2,\tau_{a_l}) - G_B(\tau_2,\tau_{b_l})]$$
$$(28)$$

and the symmetric $m \times m$ – matrix A is defined by

$$A^{-1} = \bar{T} - \frac{1}{2}B$$
$$\bar{T}_{kl} = \bar{T}_k \delta_{kl}$$
$$B_{kl} = G_B(\tau_{a_k},\tau_{a_l}) - G_B(\tau_{a_k},\tau_{b_l}) - G_B(\tau_{b_k},\tau_{a_l}) + G_B(\tau_{b_k},\tau_{b_l}).$$
$$(29)$$

Let us now return to the two-loop diagram, and try to put legs on the third line, too. To calculate insertions into the $\mathcal{D}\bar{x}$ – path integral, one might first use a transformation of variables

$$\bar{x}(\bar{\tau}) = x(\tau_a) + \frac{\bar{\tau}}{\bar{T}}[x(\tau_b) - x(\tau_a)] + \bar{y}(\bar{\tau}) \qquad (30)$$

in this path integral, and then evaluate the integral over $\mathcal{D}\bar{y}$ using the worldline Green function $G^0$ appropriate to the boundary conditions $y(0) = y(T) = 0$, which is related to $G_B$ by

$$G^0(\tau_1,\tau_2) = G_B(\tau_1,\tau_2) - G_B(\tau_1,0) - G_B(0,\tau_2) \qquad (31)$$

(note that this transformation splits off the same exponential as in eq.( 13)). It is easier, though, to use the symmetry of the diagram. If we now switch to the parametrization of fig. 2b, and fix the number of external legs on lines one and three to be $n_1$ and $n_3$ resp., carrying momenta $p_1^{(1)}, \ldots, p_{n_1}^{(1)}$ and $p_1^{(3)}, \ldots, p_{n_3}^{(3)}$, eq. ( 19) turns into

$$\prod_{a=1}^{3} \int_0^\infty dT_a e^{-m^2(T_1+T_2+T_3)} [4\pi]^{-D} [T_1 T_2 + T_1 T_3 + T_2 T_3]^{-\frac{D}{2}}$$
$$\times \prod_{i=1}^{n_1} \prod_{j=1}^{n_3} \int_0^{T_1} d\tau_i^{(1)} \int_0^{T_3} d\tau_j^{(3)} \exp\left[ \sum_{k=1}^{n_1} \sum_{l=1}^{n_3} G_{13}^{sym}(\tau_k^{(1)}, \tau_l^{(3)}) p_k^{(1)} p_l^{(3)} \right.$$
$$\left. + \frac{1}{2} \sum_{k,l=1}^{n_1} G_{11}^{sym}(\tau_k^{(1)}, \tau_l^{(1)}) p_k^{(1)} p_l^{(1)} + \frac{1}{2} \sum_{k,l=1}^{n_3} G_{33}^{sym}(\tau_k^{(3)}, \tau_l^{(3)}) p_k^{(3)} p_l^{(3)} \right]$$
(32)

where the relation between $G_{11}^{sym}, G_{33}^{sym}, G_{13}^{sym}$ and $G_B^{(1)}$ should be clear.

At this stage, it is obvious – and has been checked with the Feynman calculation for the three- and four- point cases – that the addition of further $n_2$ legs on the middle line with momenta $p_1^{(2)}, \ldots, p_{n_2}^{(2)}$ can be taken into account by adding integrations $\prod_{m=1}^{n_2} \int_0^{T_2} d\tau_m^{(2)}$, and supplementing the exponent with the terms

$$\sum_{k=1}^{n_1} \sum_{l=1}^{n_2} G_{12}^{sym}(\tau_k^{(1)}, \tau_l^{(2)}) p_k^{(1)} p_l^{(2)} + \sum_{k=1}^{n_2} \sum_{l=1}^{n_3} G_{23}^{sym}(\tau_k^{(2)}, \tau_l^{(3)}) p_k^{(2)} p_l^{(3)}$$
$$+ \frac{1}{2} \sum_{k,l=1}^{n_2} G_{22}^{sym}(\tau_k^{(2)}, \tau_l^{(2)}) p_k^{(2)} p_l^{(2)} \, ,$$
(33)

where the additional functions $G_{ij}^{sym}$ are related to the ones above by permutations. Again one may absorb the diagonal terms in the exponent into the non-diagonal ones, and ends up with

$$\begin{aligned}
\tilde{G}_{11}^{sym}(\tau_1^{(1)}, \tau_2^{(1)}) &= \Delta \mid \tau_1^{(1)} - \tau_2^{(1)} \mid \left[ (T_1 - \mid \tau_1^{(1)} - \tau_2^{(1)} \mid)(T_2 + T_3) + T_2 T_3 \right] \\
&= \mid \tau_1^{(1)} - \tau_2^{(1)} \mid - \Delta(T_2 + T_3)(\tau_1^{(1)} - \tau_2^{(1)})^2 \\
\tilde{G}_{12}^{sym}(\tau^{(1)}, \tau^{(2)}) &= \Delta \Big[ T_3(\tau^{(1)} + \tau^{(2)})[T_1 + T_2 - (\tau^{(1)} + \tau^{(2)})] \\
&\quad + \tau^{(2)}(T_2 - \tau^{(2)})T_1 + \tau^{(1)}(T_1 - \tau^{(1)})T_2 \Big] \\
&= \tau^{(1)} + \tau^{(2)} - \Delta \left[ \tau^{(1)2} T_2 + \tau^{(2)2} T_1 + (\tau^{(1)} + \tau^{(2)})^2 T_3 \right] \\
\Delta &= [T_1 T_2 + T_1 T_3 + T_2 T_3]^{-1}
\end{aligned}$$
(34)

plus permuted $\tilde{G}_{ij}$'s.

Finally, let us say a few words about generalizations. So far we have been considering scalar diagrams only, as our main object of interest was the Koba-Nielsen exponential, which does not depend on the type of particles circulating in a diagram. The inclusion of spinors now requires besides the vertex operators for the coupling to external photons, which we have already introduced in the one-loop context, also a vertex for the coupling to internal photons. The appropriate worldline expression may be seen to be

$$- \mathrm{g}\bigl(\dot{x}^\mu - 2\psi^\mu \psi_\nu \dot{\bar{x}}^\nu\bigr), \tag{35}$$

where $x^\mu$ lives on the spinor line, and $\bar{x}^\mu$ on the photon line. Corresponding expressions can be derived for other vertices occuring in standard field theory [19].

External gluons are coupled as in the one-loop case. The representation of internal gluons by worldline path integrals is, on the other hand, connected with some intricacies [5]. One-loop experience indicates that one should rather try to reduce this case to the internal scalar and fermion cases, either by using the second order formalism [2], or, more elegantly, spacetime supersymmetry [22].

The option of summing up whole classes of diagrams by "letting legs move along lines" is, of course, confined to the abelian case.

To summarize, we have shown that the worldline path integral approach to the Bern-Kosower formalism holds considerable promise as a tool for multiloop calculations. While the existence of stringlike formulas for multiloop diagrams had been proven before by more conventional methods [16, 17], the worldline path integral method has allowed us to see the *universality* of those formulas, i. e. the fact that, once a global parametrization has been chosen, the form of the resulting Koba-Nielsen exponent is independent of the number and location of the external legs. This could obviously not be seen as long as one sticks to Feynman parameters (in a sense, we have traded the uniqueness of the Feynman parameter representation for the universality of the Koba-Nielsen exponent).

We expect the resulting integral representations to become useful in a variety of contexts, ranging from multiloop QED-calculations to the two-loop effective action in nonabelian gauge theory.

*Acknowledgements:* We are grateful to O. Lechtenfeld, M. Reuter and A. A. Tseytlin for a number of enlightening discussions, and to P. Haberl for computer support. C. S. would also like to thank the Theory Group of Hannover University for hospitality during part of the time when this work was done.

# References

[1] Z. Bern and D. A. Kosower, Phys. Rev. Lett. **66** (1991) 1669;
    Z. Bern and D. A. Kosower, Nucl. Phys. **B379** (1992) 451.

[2] Z. Bern and D. C. Dunbar, Nucl. Phys. **B379** (1992) 562.

[3] Z. Bern, L. Dixon and D. A. Kosower, Phys. Rev. Lett. **70** (1993) 2677;
    Z. Bern, L. Dixon, D. C. Dunbar and D. A. Kosower, SLAC-PUB-6415 (1994).

[4] Z. Bern, D. C. Dunbar and T. Shimada, Phys. Lett. **B312** (1993) 277.

[5] M. J. Strassler, Nucl. Phys. **B385** (1992) 145.

[6] A. M. Polyakov, *Gauge Fields and Strings*, Harwood 1987.

[7] R. P. Feynman, Phys. Rev. **80** (1950) 440.

[8] E. S. Fradkin, Nucl. Phys. **76** (1966) 588.

[9] L. Brink, P. Di Vecchia and P. Howe, Nucl. Phys. **B118** (1977) 76.

[10] A. P. Balachandran, P. Salomonson, B. Skagerstam and J. Winnberg, Phys. Rev. **D15** (1977) 2308.

[11] E. S. Fradkin and A. A. Tseytlin, Phys. Lett. **B163** (1985) 123.

[12] M. J. Strassler, SLAC-PUB-5978 (1992).

[13] M. G. Schmidt and C. Schubert, Phys. Lett. **B318** (1993) 438.

[14] D. Fliegner, M. G. Schmidt and C. Schubert, HD-THEP-93-44;
     D. Fliegner, P. Haberl, M. G. Schmidt and C. Schubert, in preparation.

[15] K. Roland, Phys. Lett. **B289** (1992) 148;
     G. Cristofano, R. Marotta and K. Roland, Nucl. Phys. **B392** (1993) 345.

[16] C. S. Lam, Phys. Rev. **D 48** (1993) 873.

[17] C. S. Lam, MCGILL-93-20.

[18] C. S. Lam and J. P. Lebrun, Nuovo Cim. **59A** (1969) 397.

[19] M. G. Schmidt and C. Schubert, in preparation.

[20] L. Brink, S. Deser, B. Zumino, P. Di Vecchia and P. Howe, Phys. Lett. **B64** (1976) 435.

[21] A. O. Barut and I. H. Duru, Phys. Rep. **172** (1989) 1.

[22] Z. Bern and A. G. Morgan, UCLA-93-TEP-36.